\documentclass{article}
\usepackage{amssymb}
\usepackage{sw20lart}
\usepackage{thmsa}

\input{tcilatex}
\begin{document}

\author{Karla Weber \\
Department of Physics\\
Centre for Technological Sciences-UDESC\\
Joinville 89223-100, Santa Catarina, Brazil.\\
e-mail: karlaweberfisica@gmail.com \and J. G. Cardoso \\
Department of Mathematics\\
Centre for Technological Sciences-UDESC\\
Joinville 89223-100, Santa Catarina,\\
Brazil.\\
e-mail: dma2jgc@joinville.udesc.br
Infeld-van der Waerden spinor formalisms.}
\title{Absence of Differential Correlations Between the Wave Equations for
Upper-Lower One-Index Twistor Fields Borne by the Infeld-van der Waerden
Spinor Formalisms for General Relativity}
\date{ }
\maketitle

\begin{abstract}
It is pointed out that the wave equations for any upper-lower one-index
twistor fields which take place in the frameworks of the Infeld-van der
Waerden $\gamma \varepsilon $-formalisms must be formally the same. The only
reason for the occurrence of this result seems to be directly related to the
fact that the spinor translation of the traditional conformal Killing
equation yields twistor equations of the same form. It thus appears that the
conventional torsionless devices for keeping track in the $\gamma $%
-formalism of valences of spinor differential configurations turn out not to
be useful for sorting out the typical patterns of the equations at issue.
\end{abstract}

\section{Introduction}

Certain calculational techniques were utilized in an earlier paper [1] for
working out the twistor equation for contravariant one-index fields in
curved spacetimes. The main aim associated to the completion of the relevant
procedures was to derive one of the simplest sets of wave equations for
conformally invariant spinor fields that should presumably take place in the
frameworks of the Infeld-van der Waerden $\gamma \varepsilon $-formalisms
[2-4]. A striking feature of these wave equations is that they involve no
couplings between the twistor fields and wave functions for gravitons [5-7].
In actuality, the only coupling configurations brought about by the
techniques allowed for thereabout take up appropriate outer products
carrying the fields themselves along with some electromagnetic wave
functions for the $\gamma $-formalism [4, 5]. Loosely speaking, the
non-occurrence of $\varepsilon $-formalism couplings stems even in the case
of charged fields from the applicability of a peculiar property of partially
contracted second-order covariant derivatives of spin-tensor densities which
carry only one type of indices as well as suitable geometric attributes
[8-10]. Indeed, the electromagnetic curvature contributions that normally
enter such derivative expansions really cancel out whenever the
non-vanishing entries of the valences of the differentiated densities are
adequately related to the respective weights and antiweights [4].

The present paper just brings forward the result that the above-mentioned
wave equations possess the same form as the ones for the corresponding
lower-index fields. It shall become clear that the legitimacy of this result
rests upon the fact that the spinor translation of the classical conformal
Killing equation leads to twistor equations which must be formally the same.
Consequently, the conventional covariant devices for keeping track of
valences of spinor differential configurations in the $\gamma $-formalism
[4, 6], turn out not to be useful as regards the attainment of the full
specification of the formal patterns for the field and wave equations being
considered. We mention, in passing, that such devices had originally been
built up in connection with the derivation of a system of sourceless
gravitational and electromagnetic wave equations [5], with the pertinent
construction having crucially been based upon the implementation of the
traditional eigenvalue equations for the $\gamma $-formalism metric spinors
[2, 3]. It may be said that the motivations for elaborating upon the
situation entertained herein rely on our interest in completing the work of
Ref. [1], thereby making up appropriately the set of $\gamma \varepsilon $%
-wave equations which emerge from the curved-space version of twistor
equations for one-index fields.

The paper has been outlined as follows. In Section 2, we exhibit the twistor
field equations which are of immediate relevance to us at this stage. We
look at the twistor wave equations in Section 3, but the key remarks
concerning the lack of differential correlations between them shall be made
in Section 4. It will be convenient to employ the world-spin index notation
adhered to in Ref. [11]. In particular, the action on an index block of the
symmetry operator will be indicated by surrounding the indices singled out
with round brackets. Without any risk of confusion, we will utilize a
torsion-free operator $\nabla _{a}$ upon taking account of covariant
derivatives in each formalism. Likewise, the D'Alembertian operator for
either $\nabla _{a}$ will be written as $\square $. A horizontal bar will be
used once in Section 4 to denote the ordinary operation of complex
conjugation. Einstein's equations should thus be taken as%
\[
2\Xi _{ab}=\kappa (T_{ab}-\frac{1}{4}Tg_{ab}),\text{ }T\doteqdot
T^{ab}g_{ab}, 
\]%
where $T_{ab}$ amounts to the world version of the energy-momentum tensor of
some sources, $g_{ab}$ denotes a covariant spacetime metric tensor and $%
\kappa $ stands for the Einstein gravitational constant. By definition, the
quantity $(-2)\Xi _{ab}$ is identified with the trace-free part of the Ricci
tensor $R_{ab}$ for the Christoffel connexion of $g_{ab}$. The cosmological
constant $\lambda $ will be allowed for implicitly through the well-known
trace relation%
\[
R=4\lambda +\kappa T,\text{ }R\doteqdot R^{ab}g_{ab}. 
\]%
Our choice of sign convention for $R_{ab}$ coincides with the one made in
Ref. [11], namely,%
\[
R_{ab}\doteqdot R_{ahb}{}^{h}, 
\]%
with $R_{abc}{}^{d}$ being the corresponding Riemann tensor. We will
henceforth assume that the local world-metric signature is $(+---)$. The
calculational techniques referred to before shall be taken for granted at
the outset.

\section{Twistor equations}

The differential patterns borne by the original formulation of twistor
equations in a curved spacetime [12-14] may be thought of as arising in
either formalism from%
\begin{equation}
\nabla ^{(AA^{\prime }}K^{BB^{\prime })}=\frac{1}{4}(\nabla _{CC^{\prime
}}K^{CC^{\prime }})M^{AB}M^{A^{\prime }B^{\prime }},  \label{e1}
\end{equation}%
and\footnote{%
The symmetry operation involved in Eqs. (1) and (2) must be applied to the
index pairs.}%
\begin{equation}
\nabla _{(AA^{\prime }}K_{BB^{\prime })}=\frac{1}{4}(\nabla ^{CC^{\prime
}}K_{CC^{\prime }})M_{AB}M_{A^{\prime }B^{\prime }},  \label{e2}
\end{equation}%
where the $K$-objects amount to nothing else but the Hermitian spinor
versions of a null conformal Killing vector, and the kernel letter $M$
accordingly stands for either $\gamma $ or $\varepsilon $.

It should be emphatically observed that the genuineness of (1) and (2) as a
system of equivalent field equations lies behind a general
covariant-constancy property of the Hermitian connecting objects for both
formalisms [2, 3]. Thus, these equations can be obtained from one another on
the basis of the metric-compatibility requirements%
\begin{equation}
\nabla _{a}(M^{AB}M^{A^{\prime }B^{\prime }})=0\Leftrightarrow \nabla
_{a}(M_{AB}M_{A^{\prime }B^{\prime }})=0.  \label{e3}
\end{equation}%
Hence, by putting into effect the elementary outer-product prescription%
\begin{equation}
K^{AA^{\prime }}=\xi ^{A}\xi ^{A^{\prime }},  \label{e4}
\end{equation}%
along with its lower-index version, after accounting for some manipulations,
we get the statements%
\begin{equation}
\nabla ^{A^{\prime }(A}\xi ^{B)}=0,\text{ }\nabla _{A^{\prime }(A}\xi
_{B)}=0,  \label{e5}
\end{equation}%
which, when combined together with their complex conjugates, bring out the
typical form of twistor equations. We stress that solutions to twistor
equations are generally subject to strong consistency conditions (see, for
instance, Ref. [1]).

Either $\xi $-field of (5) bears conformal invariance [13, 14], regardless
of whether the underlying spacetime background bears conformal flatness. In
the $\gamma $-formalism, the entries of the pair $(\xi ^{A},\xi _{A})$, and
their complex conjugates, come into play as spin vectors under the action of
the Weyl gauge group of general relativity [15], whereas their $\varepsilon $%
-formalism counterparts appear as spin-vector densities of weights $%
(+1/2,-1/2)$ and antiweights $(+1/2,-1/2)$, respectively.

\section{Wave equations}

In the $\gamma $-formalism, $\xi ^{A}$ shows up [1] as a solution to the
wave equation%
\begin{equation}
(\square -\frac{R}{12})\xi ^{A}=\frac{2i}{3}\phi ^{A}{}_{B}\xi ^{B},
\label{e6}
\end{equation}%
with $\phi ^{A}{}_{B}$ denoting a wave function for Infeld-van der Waerden
photons [16-18]. In order to derive in a manifestly transparent manner the $%
\gamma $-formalism wave equation for the lower-index field $\xi _{A}$, we
initially recast the second of the statements (5) into%
\begin{equation}
2\nabla _{A^{\prime }A}\xi _{B}=\gamma _{AB}\gamma ^{LM}\nabla _{A^{\prime
}L}\xi _{M},  \label{e7}
\end{equation}%
and then operate on (7) with $\nabla _{C}^{A^{\prime }}$.\ It follows that,
calling upon the splitting [5]%
\begin{equation}
\nabla _{C}^{A^{\prime }}\nabla _{AA^{\prime }}=\frac{1}{2}\gamma
_{AC}\square -\Delta _{AC},  \label{e8}
\end{equation}%
together with the definition%
\begin{equation}
\Delta _{AC}\doteqdot -\nabla _{(A}^{A^{\prime }}\nabla _{C)A^{\prime }},
\label{e9}
\end{equation}%
and the property [4]%
\begin{equation}
\nabla _{a}(\gamma _{AB}\gamma ^{LM})=0,  \label{e10}
\end{equation}%
we arrive at%
\begin{equation}
\square \xi _{A}-\frac{2}{3}\Delta _{A}{}^{B}\xi _{B}=0.  \label{e11}
\end{equation}%
The explicit calculation of the $\Delta $-derivative of (11) gives%
\begin{equation}
\Delta _{A}{}^{B}\xi _{B}=\frac{R}{8}\xi _{A}+i\phi _{A}{}^{B}\xi _{B},
\label{e12}
\end{equation}%
whence, fitting pieces together suitably, yields%
\begin{equation}
(\square -\frac{R}{12})\xi _{A}=\frac{2i}{3}\phi _{A}{}^{B}\xi _{B}.
\label{e13}
\end{equation}

It should be evident that the equality (11) remains formally valid in the $%
\varepsilon $-formalism as well. Therefore, since the $\varepsilon $%
-formalism field $\xi _{A}$ is a covariant one-index spin-vector density of
weight $-1/2$, the $\varepsilon $-counterpart of the derivative (12) has to
be expressed as the purely gravitational contribution\footnote{%
For a similar reason, the $\varepsilon $-formalism version of (6) reads $%
(\square -\frac{R}{12})\xi ^{A}=0$. It will become manifest later in Section
4 that the relation (14) is compatible with this assertion.}%
\begin{equation}
\Delta _{A}{}^{B}\xi _{B}=\frac{R}{8}\xi _{A}.  \label{e14}
\end{equation}%
Hence, the $\varepsilon $-formalism statement corresponding to (13) must be
spelt out as%
\begin{equation}
(\square -\frac{R}{12})\xi _{A}=0.  \label{e15}
\end{equation}

\section{Concluding remarks and outlook}

The formulae shown in Section 3 supply the entire set of wave equations for
one-index conformal Killing spinors that should be tied in with the context
of the $\gamma \varepsilon $-frameworks. It is worth pointing out that the
common overall sign on the right-hand sides of (6) and (13), is due to the $%
\gamma $-formalism metric relationship between the differential
configuration (12) and%
\[
\Delta ^{A}{}_{B}\xi ^{B}=-\hspace{0.02cm}(\frac{R}{8}\xi ^{A}+i\phi
^{A}{}_{B}\xi ^{B}),
\]%
with the aforesaid relationship actually coming about when we invoke the
well-known derivatives [4]%
\[
\Delta _{AB}\gamma _{CD}=2i\phi _{AB}\gamma _{CD},\text{ }\Delta _{AB}\gamma
^{CD}=-2i\phi _{AB}\gamma ^{CD}.
\]%
What happens with regard to it is, in effect, that the pieces of those
contracted $\Delta \xi $-derivatives somehow compensate for each other while
producing the formal commonness feature of the apposite couplings through%
\[
\Delta _{A}{}^{B}\xi _{B}+\Delta {}_{AB}\xi ^{B}=2i\phi _{A}{}^{B}\xi _{B}.
\]

At first sight, one might think that a set of differential correlations
between the $\gamma $-formalism wave equations for $\xi ^{A}$ and $\xi _{A}$
could at once arise out of utilizing the devices [4, 5]%
\[
\square \xi ^{A}=\gamma ^{AB}\square \xi _{B}+(\square \gamma ^{AB})\xi
_{B}+2(\nabla ^{h}\gamma ^{AB})\nabla _{h}\xi _{B}, 
\]%
and%
\[
\square \xi _{A}=\gamma _{BA}\square \xi ^{B}+(\square \gamma _{BA})\xi
^{B}+2(\nabla ^{h}\gamma _{BA})\nabla _{h}\xi ^{B}, 
\]%
in conjunction with the eigenvalue equations [2-4]%
\[
\nabla _{a}\gamma _{AB}=i\beta _{a}\gamma _{AB},\text{ }\nabla _{a}\gamma
^{AB}=(-i\beta _{a})\gamma ^{AB}, 
\]%
and%
\[
\square \gamma ^{AB}=-\hspace{0.02cm}\Theta \gamma ^{AB},\text{ }\square
\gamma _{AB}=-\hspace{0.02cm}\overline{\Theta }\gamma _{AB}, 
\]%
where%
\[
\Theta \doteqdot \beta ^{h}\beta _{h}+i\nabla _{h}\beta ^{h}, 
\]%
and $\beta _{a}$ is a gauge-invariant real world vector. If any such
raising-lowering device were implemented in a straightforward way, then a
considerable amount of "strange" information would thereupon be brought into
the picture whilst some of the contributions involved in the intermediate
steps of the calculations that give rise to the characteristic statements%
\[
\nabla ^{(A(A^{\prime }}K^{B^{\prime })B)}=0,\text{ }\nabla _{(A(A^{\prime
}}K_{B^{\prime })B)}=0, 
\]%
would eventually be ruled out. We can conclude that any attempt at making
use of a metric prescription to recover either of (6) and (13) from the
other, would visibly carry a serious inadequacy in that the twistor
equations (5) could not be brought forth simultaneously. It is obvious that
the property we have deduced ultimately reflects the absence of index
contractions from twistor equations.

It would be worthwhile to derive the $\gamma \varepsilon $-wave equations
for twistor fields of arbitrary valences. This issue will probably be
considered further elsewhere.

\textbf{ACKNOWLEDGEMENT:}

One of us (KW) should like to acknowledge the Brazilian agency CAPES for
financial support.

\end{document}